\documentclass[pre,aps,showpacs,pretrint]{revtex4}
\usepackage{graphicx}

\newcommand{\beq}{\begin{equation}}
\newcommand{\eeq}{\end{equation}}
\newcommand{\bqn}{\begin{eqnarray}}
\newcommand{\eqn}{\end{eqnarray}}
\newcommand{\bqns}{\begin{eqnarray*}}
\newcommand{\eqns}{\end{eqnarray*}}
\newcommand{\bary}{\begin{array}}
\newcommand{\eary}{\end{array}}

\begin{document}
\title{Influences of source displacement on the features of
subwavelength imaging of a photonic crystal slab}
\author{Pi-Gang Luan, Chen-Yu Chiang, Hsiao-Yu Yeh}
\address{Wave Engineering Laboratory, Department of Optics and
Photonics, National Central University, Jhungli 320, Taiwan}

\begin{abstract}%
In this paper we study the characteristics of subwavelength
imaging of a photonic crystal (PhC) superlens under the influence
of source displacement. For square- and triangular-lattice
photonic crystal lenses, we investigate the influence of changing
the lateral position of a single point source on the imaging
uniformity and stability. We also study the effect of changing the
geometrical center of a pair of sources on the resolution of the
double-image. Both properties are found to be sensitive to the
displacement, which implies that a PhC slab cannot be treated
seriously as a flat lens. We also show that by introducing
material absorption into the dielectric cylinders of the PhC slab
and widening the lateral width of the slab, the imaging uniformity
and stability can be substantially improved. This study helps us
to clarify the underlying mechanisms of some recently found
phenomena concerning imaging instability.
\end{abstract}
 \maketitle

\section{Introduction}

Since John Pendry \cite{Pendry} proposed that a perfect lens can
in principle be designed  to overcome  the diffraction limit
\cite{XiangadnZhaowei}, issues concerning subwavelengh imaging
have become very important and popular \cite{Shen, Luan03,
prlmetalens, Aplsuperlens, WTLu, CJJP, prlsuperlens,
naturesuperlens, ZYL, SailingHe1, SailingHe2, Xwang, Belhadj, MKS,
Luan07, Suk, EDG, Ye}. A device that can focus the light from a
point source into a spot of subwavelengh width is called a
superlens. Two kinds of superlens have been studied most. The
first is a slab of left-handed metamaterial with appropriate
negative permittivity and permeability, both close to
$-1$\cite{Pendry,Shen,Luan03, prlmetalens,Aplsuperlens}. The
second is a slab of photonic crystal (PhC) with properly chosen
band structure and equal-frequency contour (EFC) \cite{CJJP,
prlsuperlens, naturesuperlens, ZYL, SailingHe1,
SailingHe2,Xwang,Belhadj, MKS, Luan07}. Both structures are
recognized by most researchers as examples of flat lens, which
means they have the advantage of having no optical axis. In fact,
there are also structures which fall between the above categories,
namely, the Mie-resonance based artificial structures operating in
the subwavelength regime \cite{Yano1,Yano2}, however, in this
paper we will focus our attention only on the previous two
categories. The transmission characteristics of the metamaterial
slab can be explained with the effective theory of the
metamaterial medium, whereas the electromagnetic behaviors of the
PhC slab can only be understood through a detailed analysis of the
PhC band structure. The comparison between them reveals that,
although these two kinds of superlens have similar functions in
focusing light and both consist of periodically arranged elements
such as resonators and dielectric rods, they are in fact belonging
to two different categories and should be distinguished carefully.
One main difference might be that for a specific operating
wavelength of subwavelength imaging, the corresponding lattice
constant in the former structure is usually smaller than in the
latter. Typically the operation wavelength of a PhC slab lens is
about 3 to 5 times of the lattice constant. This indicates that a
PhC slab is not a flat lens rigorously, and the non-uniformity of
the structure should be considered carefully because it may have
observable effects on the subwavelength imaging phenomena.

Previous studies have revealed that the surface waves decaying
away from the PhC surface usually play important role in the
subwavelength imaging \cite{MKS, Luan07}. Since surface waves are
transversely localized Bloch waves guided along a PhC surface,
they must carry the information that manifests the surface
structure (i.e., the arrangement of rods or holes and the way of
termination) of the PhC. Now, if subwwavelength images are located
near the PhC surface, then both the surface structure and the
sources position can influence the features of the images. The
influences of surface termination have already been studied in
Ref.\cite{MKS,SailingHe2}. Recently, the effects of source
displacement on the subwavelength imaging have also been found in
the studies concerning the subwavelength imaging in phononic
crystal lens \cite{Suk} and photonic quasicrystal lens \cite{EDG}.
According to these studies, researchers found that the laterally
displacement of the source can influence the image performance
substantially. In fact, similar effect in acoustic imaging system
had already been reported previously \cite{Ye}.

In this paper, we further study the influences of the source
position on the subwavelength imaging of a PhC lens. For
square-lattice and triangular-lattice PhCs consisting of
dielectric cylinders, we first investigate the influence of
changing the lateral position of a point source on the uniformity
and stability of the single image. We then study the effects of
moving the geometrical center of a pair of sources on the
resolution of the double-image. We also discuss how to improve the
imaging quality by including absorption effect into the dielectric
cylinders consisting the PhC superlens.

\section{Phenomena and discussion}

\begin{figure}
    \includegraphics[width=5in]{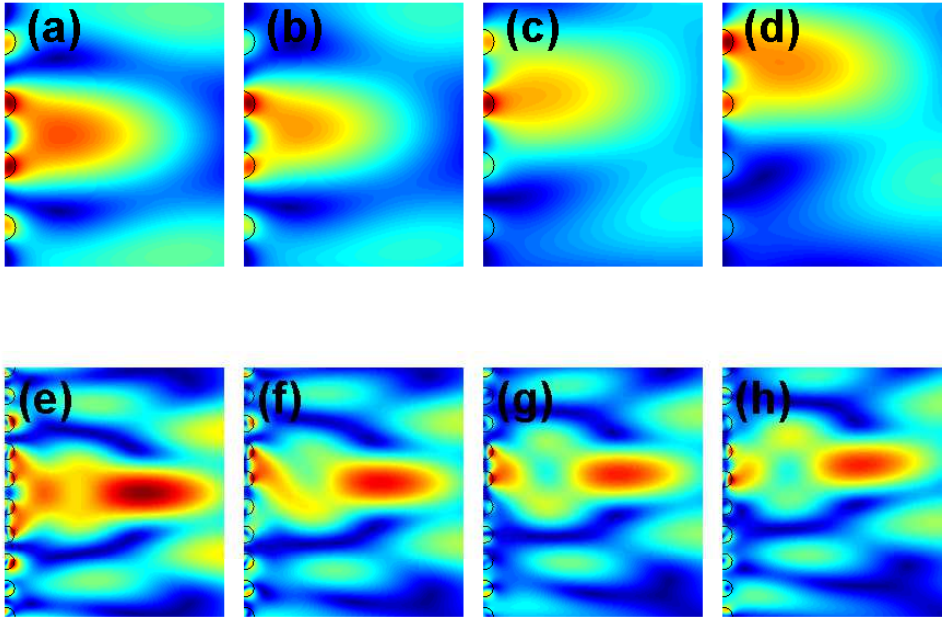}
    \caption{The image field patterns (the absolute value of the ${\bf E}$ fields) for the TM waves. Plots from
    (a) to (d) are for the square lattice cases of lateral source position: $0$,
    $\sqrt{2}a/3$,$2\sqrt{2}a/3$, and $\sqrt{2}a$. Plots from (e) to (h)
    are for the triangular lattice cases of lateral source positions: $0$, $a/3$, $2a/3$, and $a$.}
\end{figure}

We first study the imaging characteristics of a square lattice
slab consisting of dielectric cylinders in air background. In this
paper only the TM (E-polarized) wave will be studied. The
dielectric permittivity of the cylinders is $\epsilon=14$, and the
cylinder radius is $r_0=0.3a$, here $a$ is the lattice constant.
The slab has 8 layers in the $x$ direction and 21 layers in the
$y$ (lateral) direction. The orientation of the air-PhC interfaces
(slab surfaces) are $\Gamma{\rm M}$. The dimensionless frequency
$\omega a/2\pi c$ of the source is 0.192, which is in the
all-angle negative refraction frequency range \cite{CJJP}. In the
beginning, the $x$-coordinate of the source is $0.5a$ away from
the front (left) surface of the slab, located on $x$ axis ($y=0$),
which is the symmetric axis of the slab. Note that the source-slab
distance ($0.5a$) is only about one-tenth of the wavelength
($5.2a$), thus any kind of near field effect caused by the
evanescent waves may play important role in the imaging process
\cite{Luan07}. Fig 1.(a) is the electric field pattern (absolute
value) of the image behind the back surface of the slab before we
move the source. Now we keep the distance between the source and
the front surface of the slab fixed, and move the source along the
lateral direction. We notice that the image peak is not always
located on the original image plane, especially in the cases 1(b)
and 1(c). Even when the source has been moved one lateral period
of distance $\sqrt{2}a$, the image field pattern does not recover
its original form. In the second case we consider triangular
lattice PhC slab. The results of this case are shown in figure
1(e) to 1(h). Again, the dielectric cylinders are embedded in the
air background and the cylinder parameters are the same as in the
square lattice case except that now the PhC-air interfaces are
along the $\Gamma K$ direction and the dimensionless source
frequency is 0.335, in the negative refraction range of the
frequency \cite{SailingHe2}. The source is located $3.7a$ from the
front interface, longer than one wavelength ($2.99a$). This means
that the evanescent waves do not contribute much to the focused
image. In figures 1(e) to 1(h) we find that the image peak always
keeps staying on the same image plane and the symmetrical field
pattern around the peak is almost uninfluenced when the source
position has been changed.

\begin{figure}
    \includegraphics[width=5.3in]{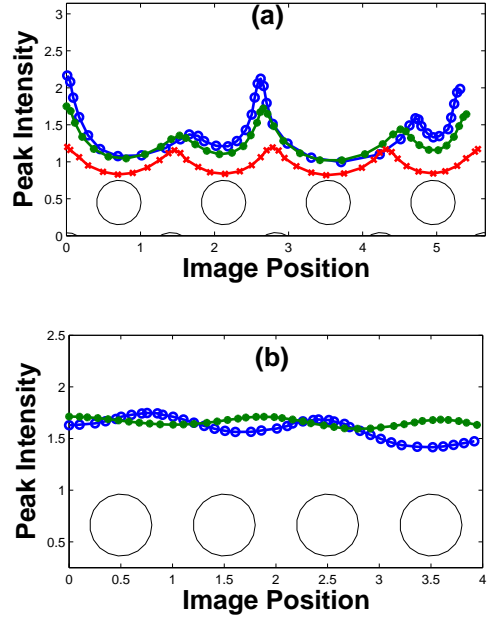}
    \caption{The image peak intensity v.s. the source position. (a) The results for the square lattice cases.
    The source changing its position continuously from the central position $y=0$ to a maximum lateral distance
    $y=4\sqrt{2}a$ is shown in the plot (blue o). The results for an absorptive slab (red x) and a
    laterally-wider slab (51 cylinders along the lateral direction) (green $*$) are also
    shown. In calculating the results of the absorptive case, we choose $\epsilon=14+0.2i$.
    (b) The results for the triangular lattice cases.
    The source changing its position continuously from $y=0$ to $y=4a$ is shown (blue o). The result for a
    wider (51 laterally arranged cylinders) slab is also shown (green $*$).}
\end{figure}
\begin{figure}
    \includegraphics[width=5in]{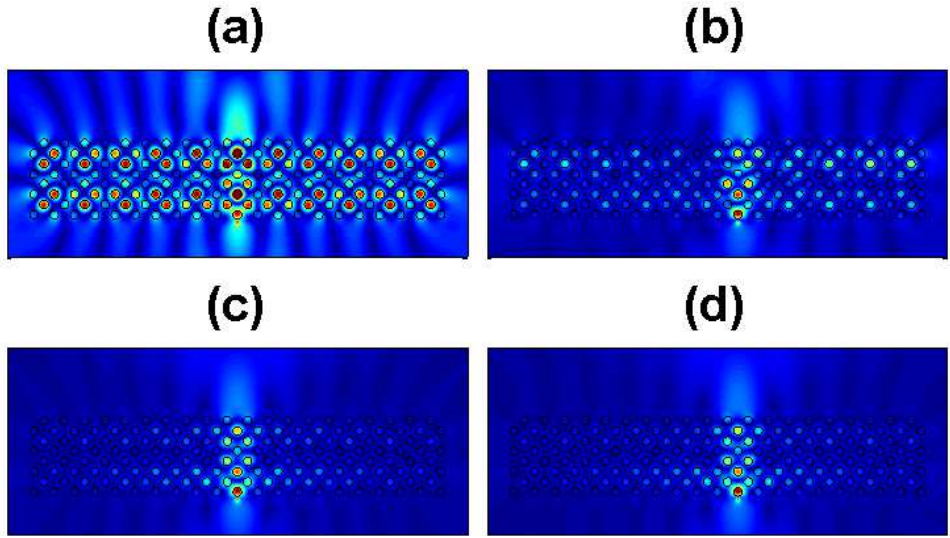}
    \caption{The field patterns (the absolute value of the ${\bf E}$ fields) for a 21-layer wide slab lens.
    Plots (a) and (b) are for the square lattice structure without source displacement and with a
    $\sqrt{2}a$ lateral displacement of the source position, respectively. Plots (c) and (d) are for
    the same structure and situations but the original permittivity of the dielectric
    cylinders are replaced by $\epsilon=14+0.2i$.}
\end{figure}
\begin{figure}
    \includegraphics[width=5in]{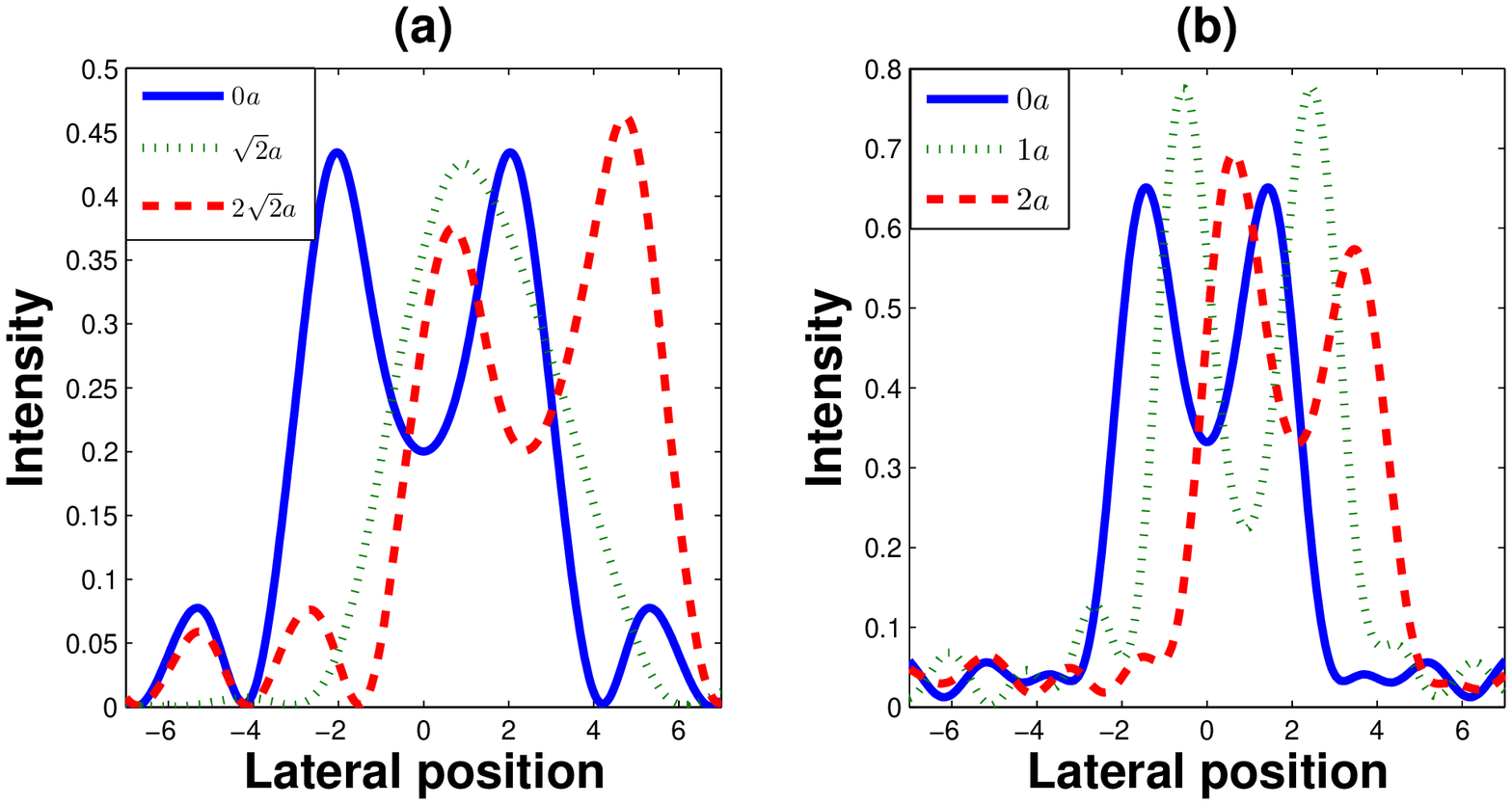}
    \caption{Intensity distribution on the image plane for the double-source case when the center of the
    source-pair is shifted 0 (solid line), one lateral period (dotted line), and two lateral periods
    (dashed line). Results in (a) are for the square lattice case and results in (b) are for the triangular
    lattice case.}
\end{figure}

When changing the source position along the lateral direction, two
interesting features are found. The first is about the periodicity
of the image strength curve and the other is about the homogeneity
of the peak strength under the source movement. For a specific
position of the source, the corresponding position and the
strength of the image peak (maximum) is recorded and designated
as, for example, an ${\rm o}$ sign. We then move the source
laterally to a nearby point and record the new position and
strength of the peak. After doing this operation many times, we
collect a lot of data about the strength and position of the image
peak. We can then fit these data with a smooth curve, and this
curve is defined as the image strength curve. Figure 2(a) shows
the results for the square lattice case having 21 cylinders along
the $y$-direction. The peak intensity with respect to the source
position reveals a inhomogeneous and non-periodic behavior (the
blue ${\rm o}$ curve). We found that the periodicity of the peak
strength can be improved a little bit by increasing the slab width
in the lateral ($y$) direction. This implies that the non-periodic
behavior is mainly caused by the finite lateral extension of the
slab. When we replace the original slab by a wider slab (51
cylinders along $y$), the periodicity becomes more obvious (the
green $*$ curve). The periodicity can be further improved by
including the absorption effect into the cylinders of the slab, as
can be easily observed (the red x curve). This can be easily
understood as follows. The absorption leads to a ``finite zone",
which defines the space region that the fields around the
cylinders are coupled together. If the zone is much smaller than
the slab width then the ``finite slab width effect" becomes
unimportant, and the periodicity recovers. This ``periodic but
inhomogeneous" behavior implies that evanescent waves still play
important role in the subwavelength imaging process even when the
absorption effect has been included. According to Fig.2(b), the
peak strength homogeneity is much better in the triangular lattice
case then in the square lattice case. However, the periodicity of
the image strength is still not obvious. The difference implies
that evanescent waves are not essential in forming the image in
the triangular lattice case.

Note that the source E field around the 2D point source is
proportional to the Hankel function $H^{(1)}_0(k|{\bf r}-{\bf
r}_s|)$, where ${\bf r}$ and ${\bf r}_s$ are the observation point
and source point, respectively, and $k=\omega/c$ is the wave
number. The Hankel function $H^{(1)}_0({\xi})$ behaves differently
for ${\xi}\ll 1$ and ${\xi}\gg 1$ (See, for example,
Ref.\cite{arfken}), thus we can define the near field zone as the
region satisfying $k|{\bf r}-{\bf r}_s|<1$. In the near field zone
the source E field has the form $A\left(\ln\left(k|{\bf r}-{\bf
r}_s|/2\right)+\gamma\right)$, whereas beyond this zone the source
field is approximated by $A'e^{ik|{\bf r}-{\bf r}_s|}/\sqrt{k|{\bf
r}-{\bf r}_s|}$. Here $A$ and $A'$ are two complex constants and
$\gamma\approx 0.5772$ is the Euler-Mascheroni constant. If we
choose the source-slab distance (which is also the shortest
distance between a cylinder and the source) as $|{\bf r}-{\bf
r}_s|$ than in the square lattice case we have $k|{\bf r}-{\bf
r}_s|=0.6032<1$, whereas in the triangular lattice case we have
$k|{\bf r}-{\bf r}_s|=7.788\gg 1$. It is interesting to note that
for the square lattice case although the nearest cylinder of the
slab is indeed within the near field zone around the source,
however, the other cylinders are outside of this zone. This
clearly explains why the image in this case is very sensitive to
the displacement of the source. For the triangular lattice case,
however, all the cylinders of the slab are outside of the near
field zone, thus the displacement of the source does not influence
the imaging much. We conclude that the superior imaging
performance of the triangular lattice PhC slab lens is a result of
the weak contribution of the near field compared to the square
lattice case.

Numerical simulation results in Fig.3 can further demonstrate the
above mentioned ``finite slab width effect" for the square lattice
structure and the ``finite zone effect" for the absorptive case.
Fig.3(a) is the field pattern for the (non-absorptive) 21-layer
wide superlens structure, and the source is located at $y=0$. When
we move the source to $y=\sqrt{2}a$, the field pattern becomes
that in Fig.3(b). It is clear that the image strength reduced and
the image pattern become asymmetrical after the displacement. The
``finite slab width effect" can be easily observed in Fig.3(b).
For the absorptive superlens, the original dielectric cylinders
are replaced by absorptive cylinders with permittivity
$\epsilon=14+0.2i$, and the results are shown in Fig.3(c) and
3.(d).  The ``finite zone effect" can be observed clearly.

Next, we consider two point sources separated by a distance which
is the smallest one to maintain distinguishable images at the
image plane. We keep this internal source separation fixed and
shift them together along the lateral direction. In the square
lattice case, the smallest distance is about $0.66\lambda$, or
$2/3$ of the wavelength. The solid curve in Fig.4(a) shows the
double-image strength on the image plane before we shift the
source-pair. After shifting the sources laterally to $y=\sqrt{2}a$
(one lateral period), the two images become indistinguishable
(dotted line), but distinguishable images reappear when we shift
the sources to $y=2\sqrt{2}a$ (dashed line). The subwavelength
double-image resolution stability in the triangular lattices is
better, as indicated in figure 4(b). Because in this case the
sources and images are far from the slab, we can ensure that
evanescent waves do not contribute much on the subwavelength
imaging in this case.

All the results discussed above refer to point sources. However,
in realistic applications, the sources must be of finite size. We
have implemented numerical simulations to explore the influence of
the source size on the imaging features. The finite size source in
the simulations is mimicked by uniformly arranging a lot of (about
31 or so) point sources within a disk of small radius. When the
source radius is smaller than one lattice constant, the
conclusions concerning the single source imaging features still
hold. For the situation of two sources, things become more
complicated, and we have not yet got conclusive results. However,
we believe this issue is worth studying because it may provide
practical knowledge for designing useful devices.

\section{Conclusion}
In this paper we discuss the image quality stability of
square-lattice and triangular-lattice PhC slab lenses under the
influence of source movement. For square lattice slab, we found
that the imaging pattern and the image peak strength are very
unstable under the source movement. Coupling between evanescent
waves is helpful in the subwavelength imaging but the image is
sensitive to the relative position between the source and the
cylinders in the front interface of the slab
\cite{Luan07,Suk,EDG,Ye}. A more ideal slab lens seems to be the
triangular lattice slab, in which evanescent wave do not play
essential role in forming the subwavelength image, and the image
quality is more stable. Thus the triangular lattice PhC slab might
be a better choice for the real applications. We also found that
the wider the slab in the lateral direction, the better the image
periodicity, and a little absorption included can help to reduce
the boundary effects of the PhC lens. We conclude that a PhC slab
cannot be treated as a flat lens rigorously. We believe that a
clearer understanding of these affecting factors on imaging can
provide some valuable references for designing new subwavelength
imaging devices.

\section*{ACKNOWLEDGEMENT} The author gratefully acknowledge
financial support from National Science Council (Grant No. NSC
98-2112-M-008-114-MY3) of the Republic of China, Taiwan.


\end{document}